\documentclass{aa}
\usepackage[dvips]{graphicx}
\usepackage{natbib}
\usepackage{txfonts}
\usepackage{supertabular}

\begin{document}

\def\ab{$\sim$}
\def\frac{$''$\hspace*{-.1cm}}
\def\deg{$^{\circ}$\hspace*{-.1cm}}
\def\min{$'$\hspace*{-.1cm}}
\def\h2{H\,{\sc ii}}
\def\hi{H\,{\sc i}}
\def\hb{H$\beta$}
\def\ha{H$\alpha$}
\def\hd{H$\delta$}
\def\heii{He\,{\sc ii}}
\def\hg{H$\gamma$}
\def\sii{[S\,{\sc ii}]}
\def\siii{[S\,{\sc iii}]}
\def\oiii{[O\,{\sc iii}]}
\def\oii{[O\,{\sc ii}]}
\def\hei{He\,{\sc i}}
\def\sm{$M_{\odot}$}
\def\slum{$L_{\odot}$}
\def\mdot{$\dot{M}$}
\def\x{$\times$}
\def\sec{s$^{-1}$}
\def\cm2{cm$^{-2}$}
\def\mcube{$^{-3}$}
\def\lam{$\lambda$}
\def\av{$A_{V}$}

\def\aiii{Al~{\sc{iii}}\ }
\def\cii{C\,{\sc{ii}}}
\def\ciii{C\,{\sc{iii}}}
\def\civ{C\,{\sc{iv}}}
\def\ci{Ca\,{\sc{ii}}}
\def\crii{Cr\,{\sc{ii}}}
\def\feii{Fe\,{\sc{ii}}}
\def\feiii{Fe\,{\sc{iii}}}
\def\hei{He\,{\sc{i}}}
\def\heii{He\,{{\sc ii}}}
\def\nii{N\,{\sc{ii}}}
\def\niii{N\,{\sc{iii}}}
\def\niv{N\,{\sc{iv}}}
\def\nv{N\,{\sc{v}}}
\def\ni{Na~{\sc{i}}\ }
\def\nei{Ne~{\sc{i}}\ }
\def\siii{S~{\sc{iii}}\ }
\def\siv{S\,{\sc{iv}}}
\def\siii{Si\,{\sc{ii}}}
\def\siiii{Si\,{\sc{iii}}}
\def\siiv{Si\,{\sc{iv}}}
\def\tiii{Ti\,{\sc{ii}}}

\def\sk{Sk\,$-71^{\circ}51$}

\title{The LMC \h2 region N\,214C and its peculiar nebular blob
\thanks{Based on observations obtained at the European Southern 
   Observatory, La Silla, Chile;  Program 69.C-0286(A) and
69.C-0286(B).}$^{,}$ 
\thanks{Tables 2 and 3 are only available in electronic form at
the CDS via anonymous ftp to \texttt{cdsarc.u-strasbg.fr} (130.79.128.5) or via
\texttt{http://cdsweb.u-strasbg.fr/cgi-bin/qcat?J/A+A/}
}
}

\offprints{Fr\'ed\'eric Meynadier, \hspace{1cm} \\Frederic.Meynadier@obspm.fr}

\date{Received 15 December 2004/ Accepted 03 March 2005}

\titlerunning{LMC N\,214C}
\authorrunning{Meynadier et al.}

\author{F. Meynadier\inst{1} \and M. Heydari-Malayeri\inst{1}
  \and Nolan R. Walborn\inst{2}
}

\institute{LERMA, Observatoire de Paris, 61 Avenue de l'Observatoire, 
F-75014 Paris, France \and Space Telescope Science Institute, 3700 San Martin
Drive, Baltimore, Maryland 21218, USA
}

\abstract{We study the Large Magellanic Cloud \h2 region N\,214C 
using imaging and spectroscopy obtained at the ESO New Technology Telescope. 
On the basis of the highest resolution images so far obtained of the 
OB association LH\,110, 
we show that the main exciting source of the \h2 region, Sk\,$-71^{\circ}51$, is in fact 
a tight cluster of massive stars  consisting of at least 6 components  
in an area \ab\,4\frac\, wide. Spectroscopic observations allow us to 
revise the spectral type of the main component (\#\,17) to 
O2 V\,((f*)) + OB, a very rare, hot type. We also classify several 
other stars associated with N\,214C and study the extinction and 
excitation characteristics of the \h2 region. 
Moreover, we obtain {\it BVR} photometry and astrometry of 
2365 stars and from the corresponding color-magnitude diagram  study the 
stellar content of N\,214C and the surrounding LH\,110. Furthermore, 
we discover a striking compact blob of ionized gas in the outer northern 
part of N\,214C. A spherical structure of \ab\,5\frac\, in radius 
(\ab\,1.3 pc), it is split into two lobes by absorbing dust 
running diametrically through its center.  We  discuss 
the possible nature of this object. \\
}

\maketitle


\section{Introduction}

Among the Large Magellanic Cloud (LMC) \h2 regions catalogued by
\citet{henize}, N\,214 is one of the southernmost, lying  
 below the bar, at a distance of \ab\,135\min\, (\ab\,2.1 kpc in 
projection) from the famous \object{30 Doradus}  (assuming a distance modulus 
of 18.6 mag \citep{groenewegen00}).
 N\,214 appears as an elongated structure,
 \ab\,15\min\,\,\x\,4\min\, (\ab\,220 pc \x\, 60 pc), composed of at
 least 8 nebular components (A to H), most of them very dim. This gas
 complex is a noteworthy region of ongoing star formation, as
 suggested by the detection of molecular emission toward components
 N\,214C and N\,214DE \citep{israel93, heikkila}. The latter molecular
 cloud is of particular interest since it is larger, more intense, and
 moreover contains several molecular species (HCN, HCO$^{+}$, CS,
 etc.)  tracing high-density regions \citep{chin}. In fact the \hi\,
 column density toward N\,214 is one of the largest in the whole LMC
\citep{israel97}. N\,214 also harbors the OB association  \object{LH 110} \citep{lh},
a study of which was presented by \citet{oey}. \\

Few works have been devoted to the \h2 regions constituting N\,214
and its associated stellar populations. The present paper deals with
the brightest component N\,214C, also called \object{NGC 2103} or
\object{DEM L 293}
\citep{dem}. We use optical imaging and spectroscopy in order to 
study the \h2 region as well as its associated stars. N\,214C is quite
attractive since it hosts Sk\,$-71^{\circ}51$, a very hot and massive star previously 
classified O3 V((f*)) \citep{walborn02}, which our present work
shows  to be even earlier.  The most massive stars being
particularly rare, and since we do not know yet how they form, N\,214C
provides an excellent opportunity for studying the formation site of
such a star, including its related stellar populations.  \\

\section{Observations and data reduction}
\subsection{Sub-arcsecond imaging and photometry}

N\,214C was observed on 28 September 2002 using the
ESO New Technology Telescope (NTT) equipped with the active optics and
the SUperb Seeing Imager (SuSI2). The detector was made up of two CCD
chips, identified as ESO \#45 and \#46.  The two resulting frames were 
automatically combined in a single FITS file, whereas the space
between the two chips was ``filled'' with some overscan columns so that
the respective geometry of the two chips was approximatively
preserved. The gap between the chips corresponds to \ab\,100 real CCD
pixels, or \ab\,8\frac.  The file format was 4288\,\x\,4096
pixels, and the measured pixel size  0\frac.085 on the sky. Each
chip of the mosaic covered a field of 5\min.5\,\x\, 2\min.7.  See the
ESO manual SuSI2 for more technical information 
(LSO-MAN-ESO-40100-0002/1.9).\\

Nebular imaging was carried out using the narrow band filters centered
on the emission lines \ha\, (ESO \#884), \hb\, (\#881), and
\oiii\,(\#882) with basic exposures of 300 sec; for the first two
filters 2 exposures and in the latter case 4 exposures were obtained.
The image quality was quite good during the night, as represented 
by a seeing of\, 0\frac.5--0\frac.8. \\
 
Photometry was performed in the {\it BVR} Bessell system using the filters
ESO \#811 ($B$), \#812 ($V$), and \#813 ($R$).  We were
particularly careful to keep most of the brightest stars in the field
under the detector's saturation level in order to get high quality 
Point Spread Functions (PSF). This led us to adopt unit
exposure times of 60 sec for $B$ and $V$ and 30 sec for $R$
respectively. The exposures for each filter were repeated 4 times
using ditherings of 5\frac\,--10\frac\, for bad pixel rejection. \\ 

Seven standard stars, belonging to two Landolt
photometric groups (SA 92 and T Phe) were observed at four different
airmasses. This led to the determination of the photometry coefficients  
and zero-points using the \texttt{photcal} package under
\texttt{iraf}. Those coefficients are in good agreement with the
indicative values which are displayed on the SuSI2 web page. \\

The offset between standard stars' wide-aperture photometry and field
stars' PSF-fitting photometry was calculated as follows: starting from
one of the flat-fielded frames, we subtracted all stars except
the ones used for determining the PSF with the
\texttt{daophot.substar} procedure, using our preliminary DAOPHOT
photometry and the corresponding PSF. This leads to a frame with only
a few bright, isolated stars plus residues from the subtraction. We then
performed both aperture and PSF-fitting photometry on those stars,
using the same aperture we used for standard stars. Finally, we
compared the results and eliminated deviant measurements (occasional bad
PSF subtraction and strongly variable nebular background). This results in
aperture corrections of 0.02, 0.04 and 0.03 mag in $B$, $V$ and
$R$ respectively. \\

During the photometry process, some slight discrepancies between the
intensity of the frames were found: this effect was considered to be
the consequence of episodic variations of sky transparency by 7\% at
most. In order
not to introduce a systematic underestimation of the star magnitudes 
when averaging the frames, we decided to perform photometry
on each individual frame with \texttt{daophot}. 
Then, for each star, we computed the
magnitude differences from  one frame to another, and deduced the mean
magnitude shift between each frame. Choosing the brightest ones as the
reference for each filter, we multiplied the three others by the appropriate 
correction factor ($10^{-0.4\,.\,\Delta m}$, with $\Delta m$
being the (negative) mean difference of magnitude between the brightest 
frame and the current frame) and performed another run of
photometry. By cross-correlating the positions of the sources in the
various photometry files, we obtained the mean magnitude (average
of the 4 magnitudes) and a decent estimator of the uncertainty on this
magnitude (difference between maximum and minimum magnitudes, the
sample being too small for $\sigma$ to be significant). The
mean uncertainties, much larger and probably more meaningful than the
\texttt{daophot} internal errors, are reported in Table\,\ref{errors}.
Finally, the process yielded the photometry of 2321 stars in all 
three filters (online Table \ref{tab:photometrie}). \\ 

\begin{figure*}
\begin{center}
\includegraphics[width = \linewidth]{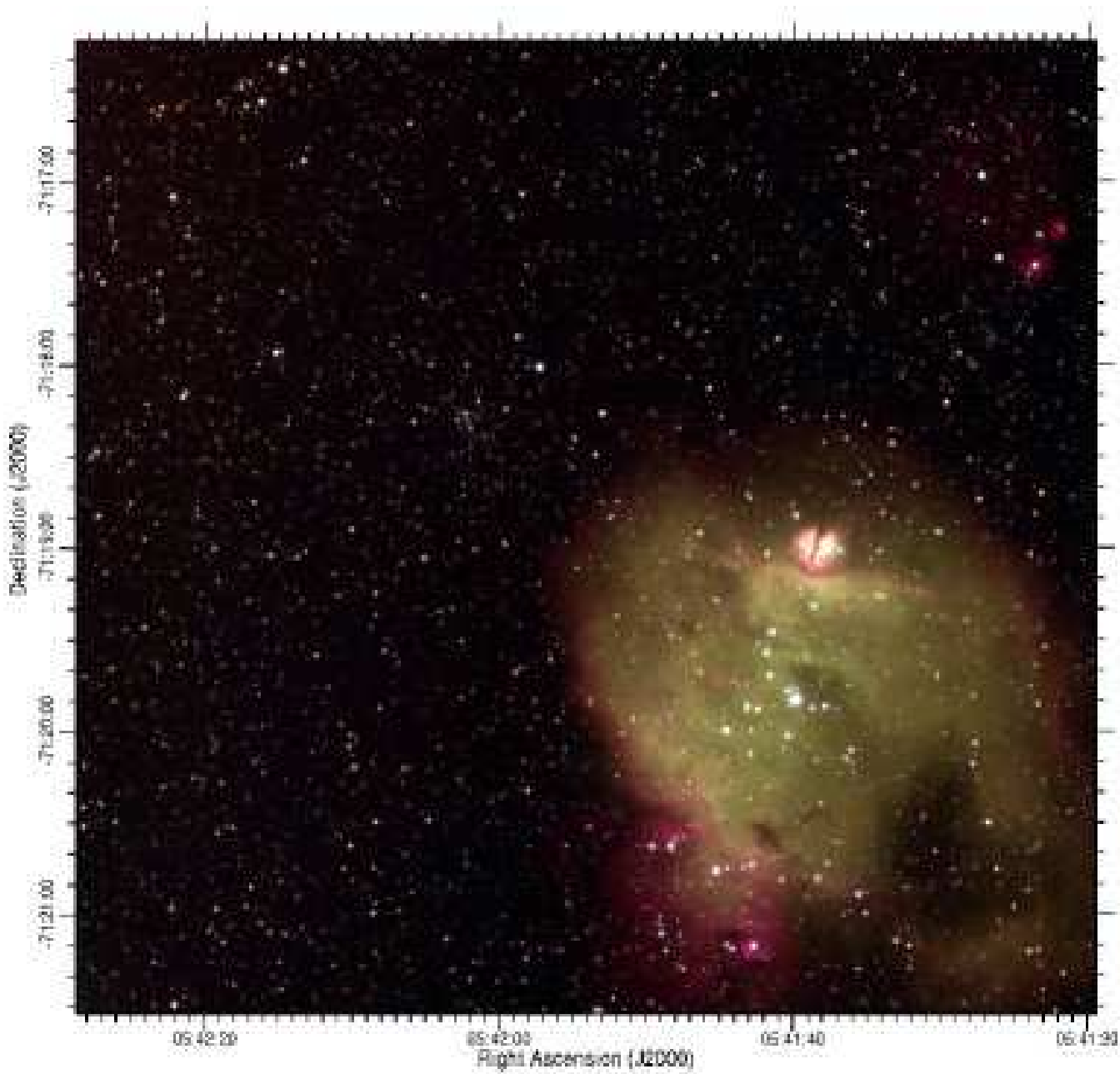}
\caption{A composite  six-color image of the LMC \h2 region N\,214C and 
the OB association LH\,110 in which it lies. The image, 
taken with the ESO NTT/SuSI2, results from the co-addition of narrow
and broad-band filters $R$ and \ha\ (red), $V$ and \oiii\
(green), and $B$ and \hb\ (blue). 
The field size is 364\frac\,\,\x\,348\frac\, corresponding to 
91\,\x\,87 pc. North is up and east to the left. 
The brightest star, situated toward the middle of the nebula, 
is the Sk\,$-71^{\circ}51$\, cluster. The striking compact \h2 blob lies  
\ab\,60\frac\, (\ab\,15 pc) north of Sk\,$-71^{\circ}51$. 
See Fig.\,\ref{fig:anonymous} for identifications of the 
compact nebulae lying north-west of N\,214C. 
\label{fig:global} } 
\end{center}
\end{figure*}

\begin{figure*}
\centering\includegraphics[width=\linewidth]{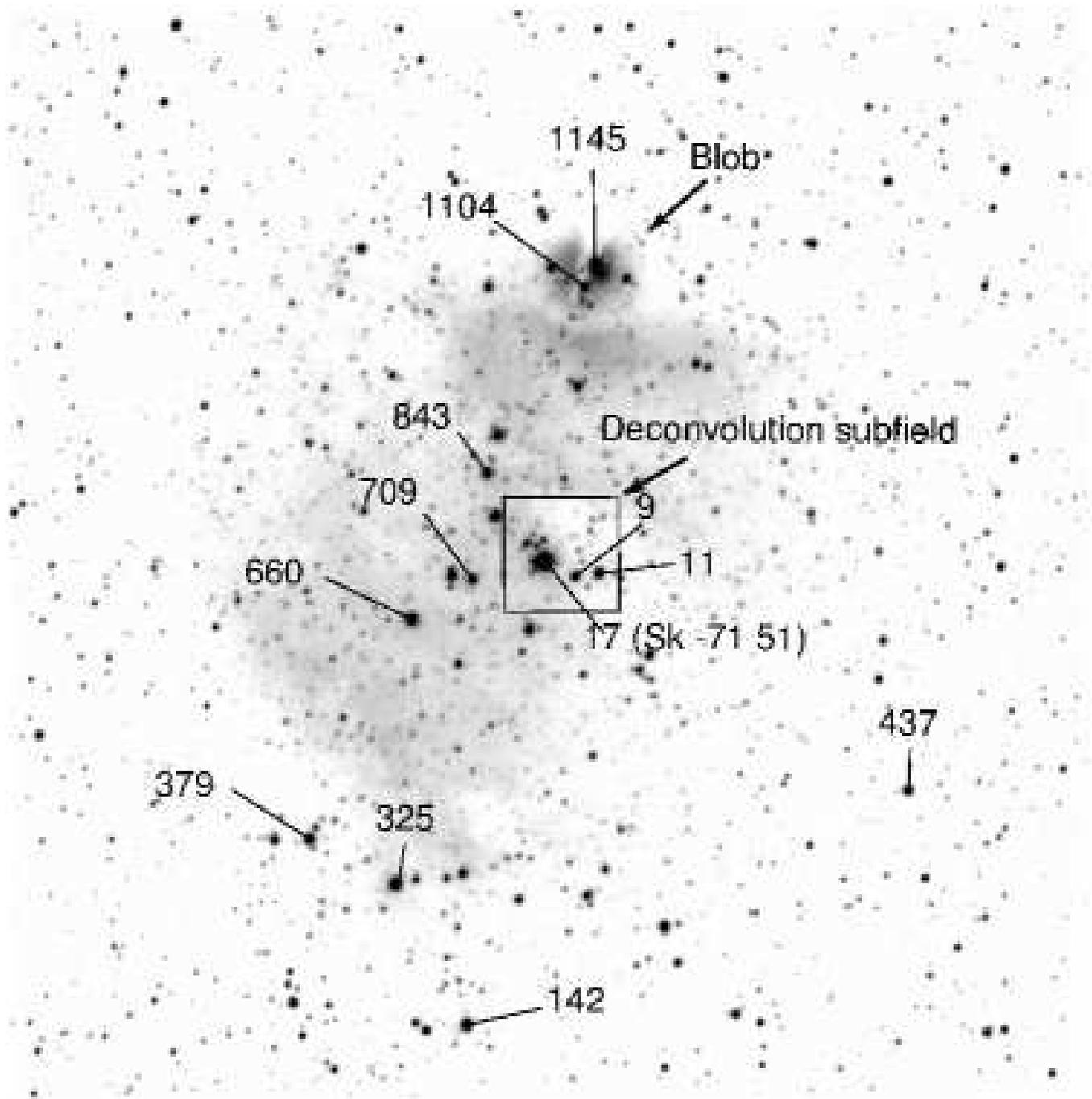}
\caption{The \h2\, region N\,214C through the broad-band filter $V$ 
obtained using ESO NTT equipped with SuSI2. North is up and east to the 
left. Field size: 206\frac\,\,\x\,208\frac\, (\ab\,52\,\x\,52\,pc). 
The stars for which spectroscopy has been 
obtained are labelled, as well as some noteworthy stars
of the sample indicated in Fig.\,\ref{col-mag}. The subfield 
(21\frac.8$^{2}$) on which the deconvolution has been carried out is 
also shown.  }
\label{fig:schema}
\end{figure*}

\begin{figure*}
  \includegraphics[width = .9\linewidth]{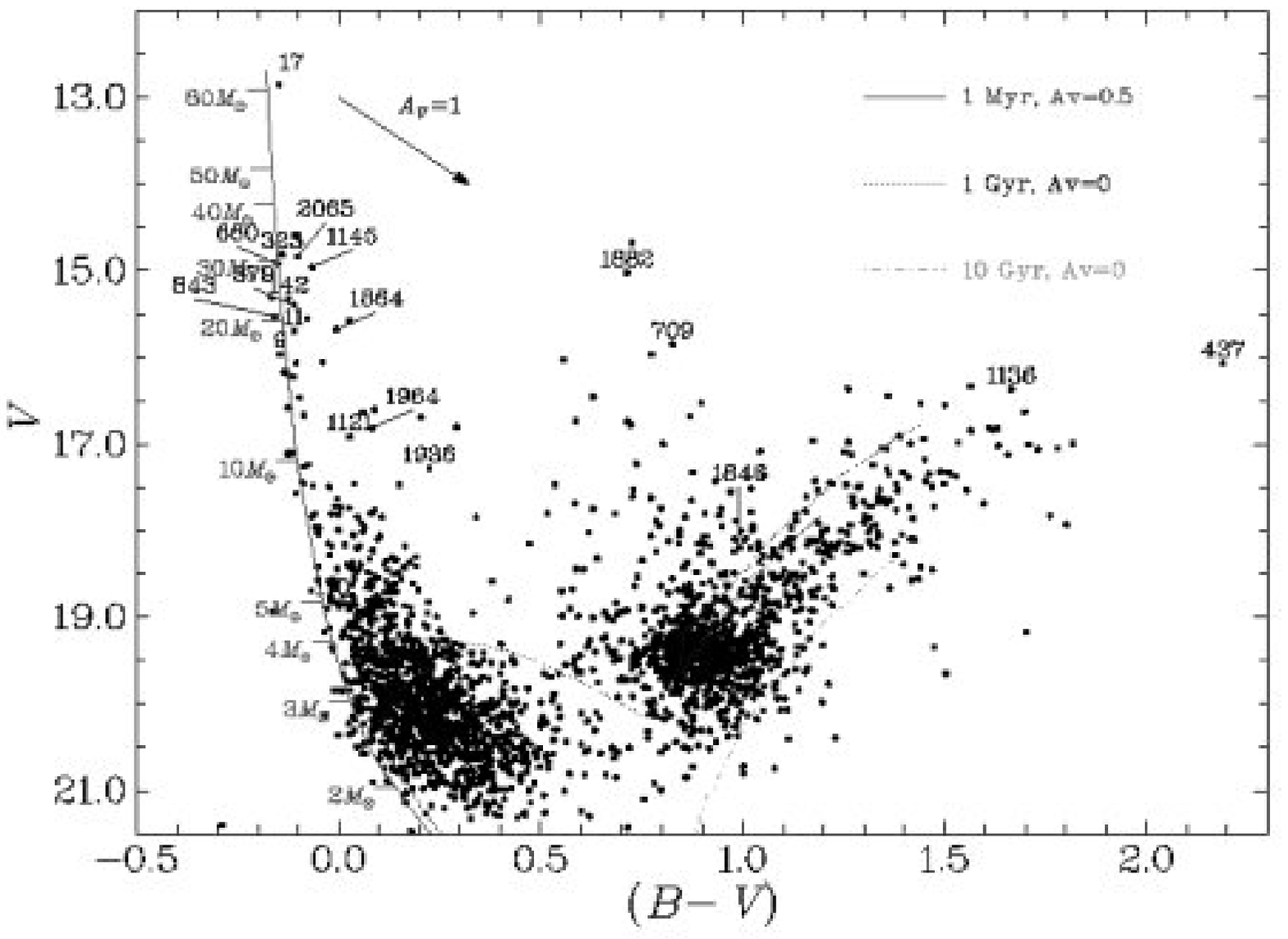}
\caption{
Color-magnitude, {\it V} versus {\it B -- V}, diagram for the
2341 stars observed  toward LMC N\,214C. 
Three isochrones are shown, 1 Myr with $A_V = 0.5$ mag (red curve), 
1 Gyr (dotted blue), and 10 Gyr (dashed-dotted green), computed for a
metallicity of Z = 0.008 \citep{lejeune01} and a distance modulus of 
18.6 mag.  A number of masses between 2 and 80\,\sm\, are indicated for the 
1 Myr isochrone.  The arrow indicates the reddening shift for 
$A_{V}$\,=\,1 mag.  The numbers refer to the stellar identifications
presented in Fig.\,\ref{fig:schema}.  
\label{col-mag}
}
\end{figure*}

A composite, six-color image of the whole N\,214C field is displayed in 
Fig.\,\ref{fig:global}, while Fig.\,\ref{fig:schema} shows a subfield in 
$V$, where the most conspicuous stars are labelled. 
The corresponding color-magnitude diagram is displayed in 
Fig.\,\ref{col-mag}, and discussed in Sect. 5.

\begin{table}[ht]
\setcounter{table}{0}
\caption{Mean overall photometric errors}
\begin{center}
\begin{tabular}{cccc}

\hline
mag & B & V & R\\
\hline
14 & 0.007 & 0.014 & 0.018\\
15 & 0.012 & 0.016 & 0.021\\
16 & 0.014 & 0.016 & 0.022\\
17 & 0.019 & 0.025 & 0.030\\
18 & 0.029 & 0.031 & 0.040\\
19 & 0.044 & 0.047 & 0.066\\
20 & 0.078 & 0.082 & 0.107\\
21 & 0.149 & 0.144 & 0.222\\
\hline
\end{tabular}
\label{errors}
\end{center}
\end{table}

\subsection{Image restoration}

\object{Sk -71 51} , the most prominent star of the field, lies in a
crowded area. In order to obtain reliable photometry for this star,
we performed a deconvolution on a subfield centered on the star. 
We used the MCS deconvolution 
algorithm, originally proposed by \citet{Magain98}.  
The strength of this code is that it takes the
finite resolution of CCDs into account and ensures that the Shannon
sampling theorem is not violated during the deconvolution
process: in particular the final PSF is chosen so that it remains
well sampled by the pixels (fwhm $\geq$ 2 pixels). 
This allows accurate photometry and
astrometry with minimum deconvolution artefacts 
for any blended sources, provided that the PSF of the
original image is adequately oversampled and that the S/N ratio of the
sources to be deconvolved remains high enough. 
We had successfully used this algorithm, in its original
implementation, for one of our previous works \citep{hmmw03}. The
present work makes use of an improved version of the MCS algorithm,
allowing simultaneous PSF determination and photometry in dense
stellar fields \citep{magain04}.\\

\begin{figure*}
\includegraphics[width=8.5cm]{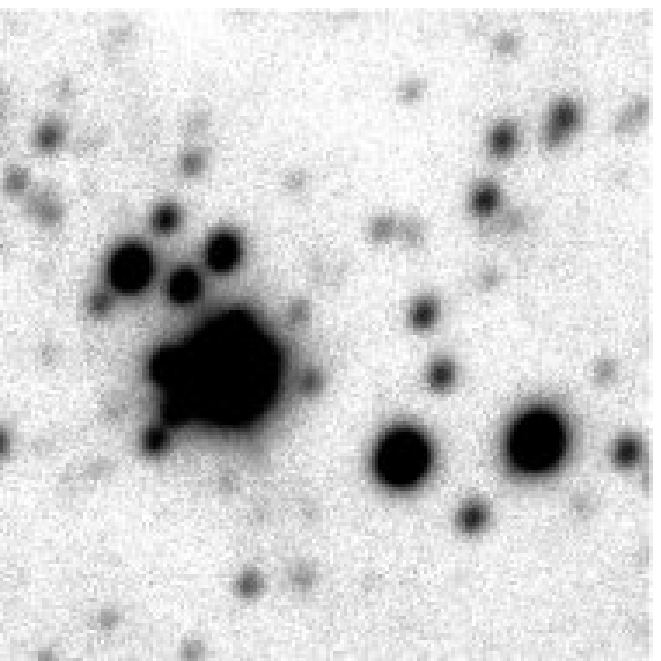}
\hfill
\includegraphics[width=8.5cm]{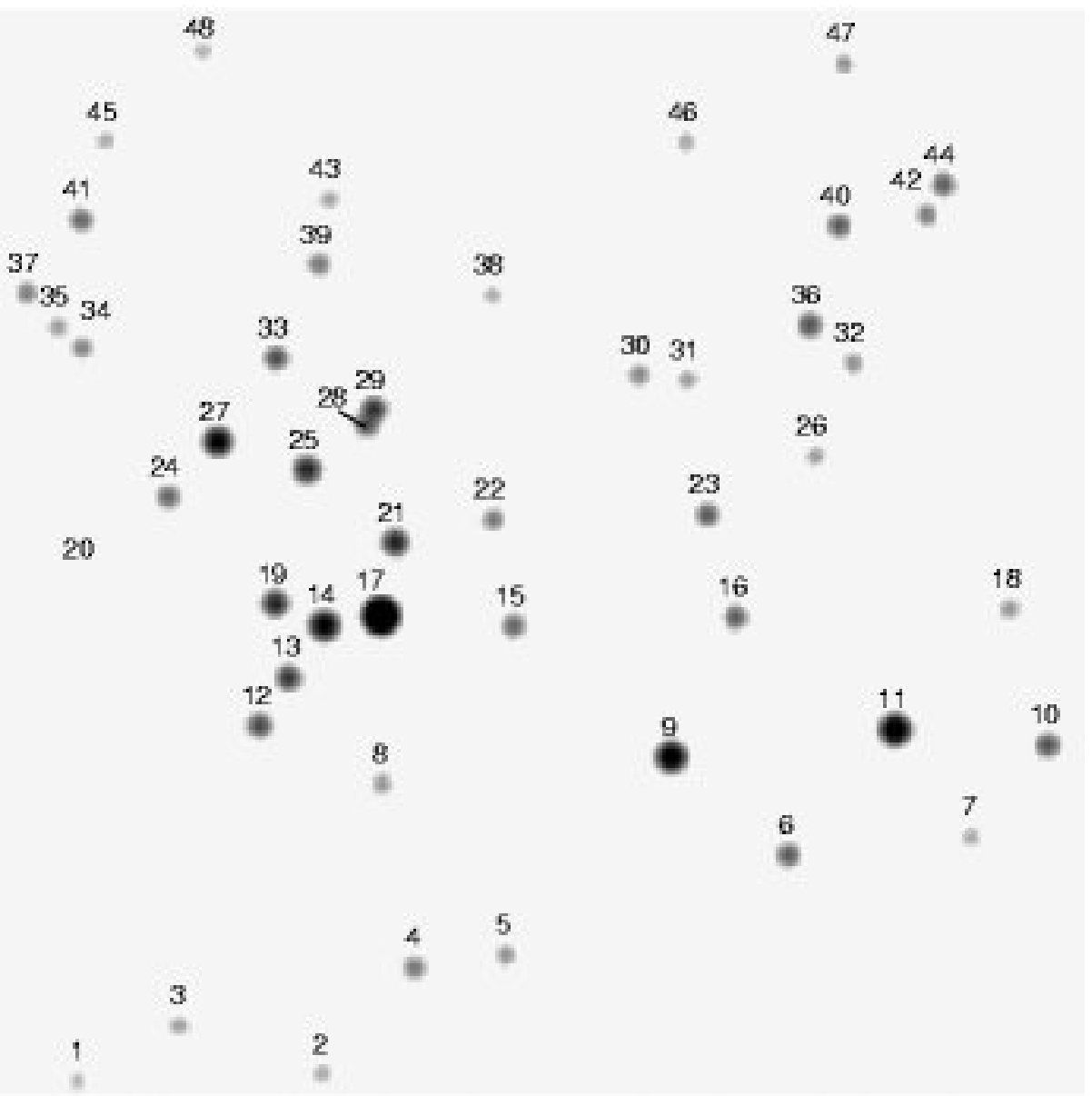}
\caption{The deconvolution subfield  around the hot star Sk\,$-71^{\circ}51$\, in $V$ 
(see Fig.\,\ref{fig:schema}).  
{\bf Left:} single frame after subtraction of the nebular background, 
exposure time 60 sec, seeing about 8.5 pixels (0\frac.72). {\bf Right:} 
the same field after deconvolution. The resulting {\sc fwhm} of the sources 
is 3 pixels,  or 0\frac.25 on the sky. 
Field size 21\frac.7$^{2}$ (\ab\,5.4$^{2}$ pc).  North is up and
east to the left. 
\label{fig:deconvolution}  
}
\end{figure*}

The original data are a set of 12 frames, 256$^{2}$  pixels each 
(21\frac.8\,$^{2}$  on the sky) 
extracted from the $B$, $V$ and $R$ individual flat-fielded frames described
earlier. Unfortunately 2 of the 4 $V$ frames were found to be 
problematic: one because of a meteoric trace close to faint sources, and the
other because of saturation of a few pixels. We conservatively chose
not to consider those two frames for deconvolution, which reduces the 
set to: 4 $B$ frames, 2 $V$ frames, and 4 $R$ frames. 
After subtraction of the nebular background using \texttt{sextractor}
\citep{bertin96}, the 10 frames have been processed 
simultaneously: the astrometry was assumed to be identical on
each frame, with a small relative shift. Finally the output consists 
of 10 photometry values and a pair of coordinates for each source: 
the mean value in each filter is computed afterwards, the spread of 
values giving an estimate of the uncertainty on instrumental magnitudes  
with this method.  As a result, the subfield is resolved into 48 stars, 
labelled from 1 to 48 (Fig.\,\ref{fig:deconvolution}), 
the photometry of 44 among them is given in 
online Table\,\ref{tab:deconvolution}.

\subsection{Spectroscopy}

The EMMI spectrograph attached to the ESO NTT telescope was used on 28
and 30 September 2002 to obtain several long slit spectra of the
stars.  The grating was \#\,12 centered on 4350\,\AA\, (BLMRD mode)
and the detector was a Tektronix CCD TK1034 with 1024$^{2}$ pixels of
size 24 $\mu$m.  The covered wavelength range was 3810--4740\,\AA\, 
and the dispersion 38\,\AA\,mm$^{-1}$, giving {\sc fwhm} 
resolutions of $2.70\pm0.10$  
pixels or $2.48\pm0.13$\,\AA\, for a 1\frac.0 slit. At each position
we first took a short 5 min exposure followed by one or two longer 15
min exposures. The instrument response was derived from 
observations of the calibration stars  LTT 7379, LTT 6248, and  LTT 7987.\\

The seeing conditions were very good, varying 
between 0\frac.4 and 0\frac.5.
The identification of the stars along the
slits was based on monitor sketches drawn during the observations. 
However, in order to avoid  ambiguous identifications more especially 
in crowded regions, we also worked out a small  \texttt{IRAF} task, 
using the position angle information in the FITS headers. 
First, each spectrum
was integrated along the dispersion axis, the result being stored into
a two-pixel wide strip, which is close to the actual size of the slit.
Then, the position angle and the pixel-arcsec correspondence were used
to calculate the rotation matrix for the World Coordinate System (WCS).
This allowed creation of a slit chart, an $\alpha$-$\delta$ calibrated
two-dimensional image containing accurate slit orientations.
Displaying simultaneously the slit chart beside the SuSI images and
using the WCS correlations it was possible to accurately check the
identity of the star in the slit. \\

EMMI was also used to obtain nebular spectra with grating 
\#\,8 (4550-6650\,\AA) in the REMD mode. The detector was  CCD \#\,63, 
MIT/LL, 2048\,\x\,4096 pixels of  15$^{2}\,\mu$m$^{2}$ each.   
A number of spectra were obtained with the slit set in 
east-west and north-south orientations using a basic exposure time 
of 300 sec repeated several times. These spectra were used to calibrate 
the line ratio maps \ha/\hb\, and \oiii/\hb\, which are 
based on imaging (See Sect. 3.2 and 3.3). \\

The stellar spectra were extracted using the \texttt{specres} iraf
task \citep{specres}, part of the \texttt{stecf} iraf package 
(\texttt{http://www.stecf.org/software/stecf-iraf/}). This procedure
uses an iterative technique involving Lucy-Richardson restoration to
avoid contamination from the inhomogeneous nebular background.

\section{Results}

\subsection{General morphology}

N\,214C is the most prominent \ha\, emission nebula of the OB association 
LH\,110 (Fig.\,\ref{fig:global}). It extends over 
\ab\,3\min.5 (\ab\,52 pc) along the north-south direction while 
its mean size along  east-west  is less than 2\min.5 (38 pc), 
becoming  narrower in its southern part. 
The bulk of stars making up LH\,110 is visible in Fig.\,\ref{fig:global}, 
and in particular a tight cluster can be seen around 
$\alpha=$ 05:42:02,\, $\delta=-$71:18:20.  \\

At the center of the nebula lies Sk\,$-71^{\circ}51$, the region's brightest and 
hottest star \citep{walborn02}.  At a distance of
\ab\,15\frac\, north of Sk\,$-71^{\circ}51$\, runs a long arc of
shocked gas with a curvature pointing to the star.  There are a
dozen  less bright stars scattered across the nebula and mainly around
Sk\,$-71^{\circ}51$. The green color in the composite image (Fig.\,\ref{fig:global}) 
covering the bulk of the \h2 region comes from the high excitation 
forbidden line \oiii\,\lam\,5007.   
Moreover, several fine, filamentary structures and fine 
absorption pillars and features are visible. The eastern or more 
particularly the south-eastern boundaries 
of the \h2 region exhibit a rim structure, evoking a cavity filled 
with ionized gas. There is however a nebular extension toward 
the south-east 
where three relatively bright stars are present 
(\#\,325, \#\,379, \#\,142) and the emission is mostly due to \ha. 
We also see a relatively large loop of 
ionized gas which protrudes from the north-west and bends to join  
the main body of the nebula at its southern part. \\

\begin{figure}[h]
\includegraphics[width =\linewidth]{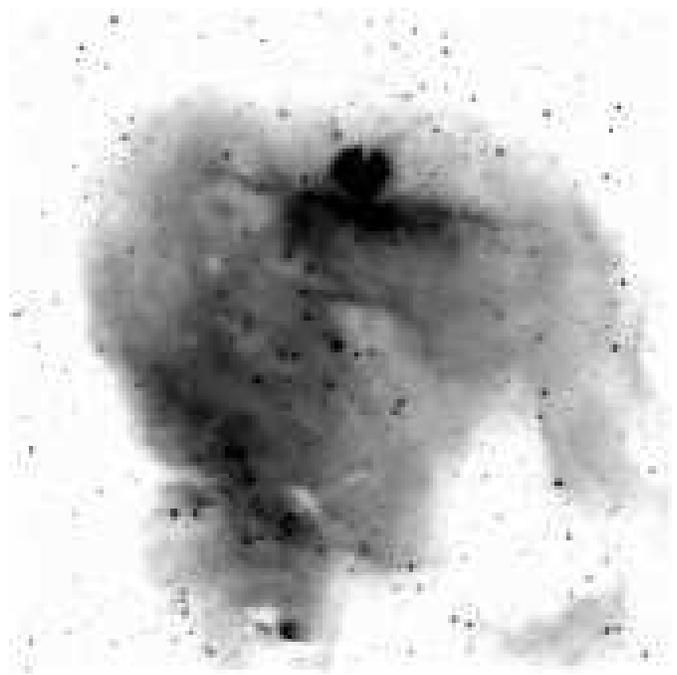}
\caption{An \ha\, image with a deep intensity cut displaying    
the northern and southern-eastern ridges. Field size the  
same as in Fig.\,\ref{fig:schema}.
\label{fig:deep}
}
\end{figure}

An \ha\, image with much deeper cut Fig.\,\ref{fig:deep} shows 
a remarkable symmetry of the nebular structures with respect to 
Sk\,$-71^{\circ}51$\, as far as the bright ridges to the north 
and south are concerned. The \h2 region appears to 
be ``breaking out'', i.e. encountering lower densities, to the 
east and west. There may be dense clouds north and south. \\

\begin{figure*}
\begin{center}
\includegraphics[width = 12cm]{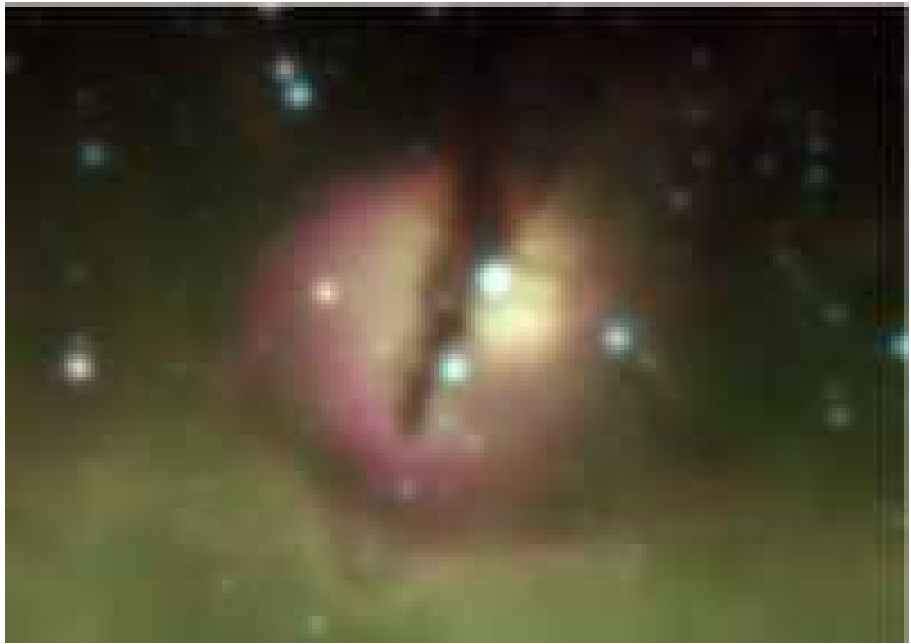}
\caption{A composite color image of the nebular blob lying 
\ab\,60\frac\, (15 pc) north of the Sk\,$-71^{\circ}51$\, cluster. Based on individual images 
taken through narrow-band filters \ha\, (red), \oiii\, (green) and  \hb\, 
(blue). Size 513\,\x\,363 pixels, 43\frac\,\x\,31\frac\,  on the sky, 
corresponding to \ab\,11\,\x\,8 pc. North is up and east to the left. 
The brightest star north of the emission peak is \#\,1145 and the one 
south of the peak is \#\,1104. The faint star situated toward the center 
of the blob is \#\,1132. The two relatively bright stars in the
eastern and western lobes are \#\,1136 and \#\,1121 respectively (see
also Table \ref{tab:classification}).
\label{blob}  
}
\end{center}
\end{figure*}

A remarkable feature of N\,214C is the presence of a globular blob of
ionized gas at \ab\,60\frac\, (\ab\,15 pc in projection) north of Sk\,$-71^{\circ}51$\, 
to which attention was drawn by \citet{walborn02}. It appears
as a sphere of radius \ab\,5\frac\, (\ab\,1.3 pc) split into two 
lobes by a dust lane which runs  
along an almost north-south direction (Fig.\,\ref{blob}). 
The absorption lane 
is \ab\,1\frac\, wide in its southern part and
becomes larger 
toward the north, reaching a width of \ab\,3\frac.5 (\ab\,0.9 pc). The
western lobe has a bright emission peak  not only in the Balmer 
\ha\, and \hb\, recombination lines, but also in the higher excitation 
\oiii\,\lam\,5007 line. Several stars are seen toward the face of the 
\h2 blob, the brightest one (\#\,1145) appears to lie \ab\,2\frac.4 
(\ab\,0.6  pc) north of the peak zone at the border of the dark lane. 
We see another conspicuous star (\#\,1104) at \ab\,3\frac.9 (\ab\,1 pc) 
south of the emission peak. And there is also a faint star (\#\,1132) 
lying apparently toward the center  of the blob. 
More generally, the blob seems to be 
placed on a ridge of ionized gas. The ridge follows the structure of 
the blob implying a possible interaction, and runs on both sides 
of the blob over \ab\,80\frac\, on the sky, corresponding to 
\ab\,20\,pc in projection.  \\

\begin{figure*}
\begin{center}
\includegraphics[width = 10cm]{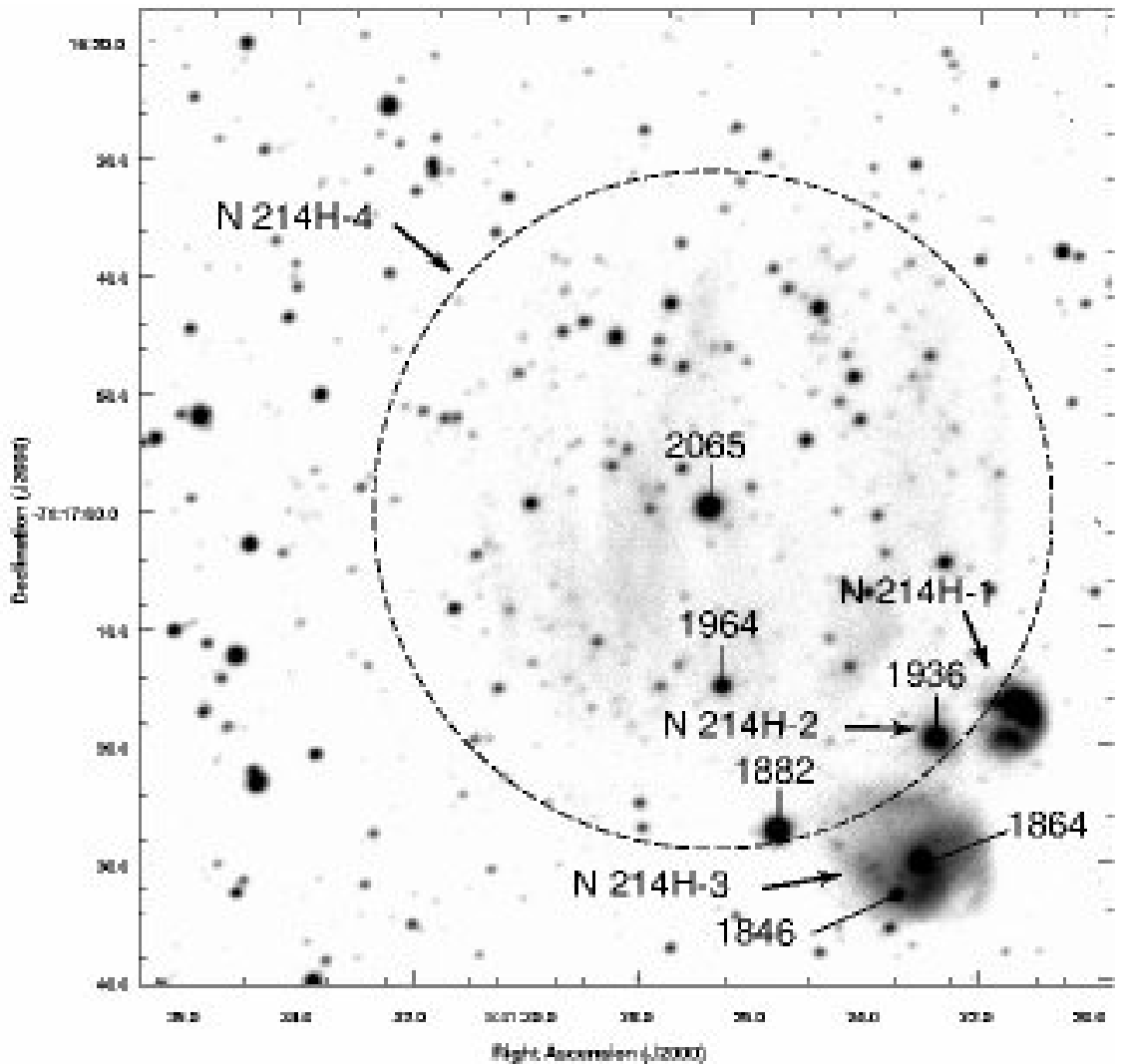}
\caption{\ha\, image of the anonymous compact nebulae, labelled N\,214H-1, 2, 
and 3 and the diffuse N\,214H-4, all lying 
north-west of N\,214C. Field size
\ab\,88\frac\,\x\,88\frac\, (\ab\,22\,\x\,22 pc).  North is up and
east to the left. See Table\,\ref{tab:classification} for more information on 
the labelled stars.
\label{fig:anonymous}  
}
\end{center}
\end{figure*}

The observations also resolve the \object{LHA 120-N 214H} 
component which lies north-west of 
N\,214C at a distance of \ab\,130\frac\, (\ab\,33 pc) from the blob
(Fig.\,\ref{fig:anonymous}). We show that N\,214H consists in fact of four 
unknown, compact \h2 regions hosting exciting stars and a diffuse 
nebula surrounding star \#\,2065. Tentatively, we call these anonymous 
nebulae N\,214H-1, 2, 3, and 4. N\,214H-1 has a structure identical to 
that of  the blob, in miniature!

\subsection{Extinction}

Fig.\,\ref{o3hb} displays the extinction map of the N\,214C region
obtained from the Balmer decrement \ha\,/\hb\, ratio. 
The most extincted part of the nebula is 
the absorption lane where the ratio reaches a value of 
\ab\,10 corresponding to \av\,\ab\,3.5  mag. It should be underlined 
that this estimate is likely a lower limit because the \ha\,/\hb\, 
ratio samples outer zones. Moreover, the Balmer ratio is not a good 
extinction indicator in this particular case since the exciting star 
lies above the lane between the blob and the observer. Outside the lane, 
the ratio is on average
\ab\,5.5 (\av\,=\,1.8 mag) and even reaches 6.2 (\av\,=\,2.2 mag)
toward the bright peak of the western lobe. The ridge on which 
the blob apparently lies also has a relatively high ratio reaching sometimes 
\ab\,4.7 (\av\,=\,1.4 mag).
Another notable reddened area is the south-eastern border of N\,214C
where the ratio fluctuates around \ab\,4 (\av\,=\,0.9 mag) and reaches
as high as \ab\,4.3 (\av\,=\, 1.1 mag). The ratio is significantly
smaller and almost uniform toward other areas of the \h2 region and
attains a value of \ab\,3 (\av\,=\,0.1 mag) in the area surrounding
the hot star Sk\,$-71^{\circ}51$. The external loop structure also has a remarkable 
reddening, in average 3.5, corresponding to \av\,=\,0.6 mag. 
Note that previous extinction estimates toward
N\,214C obtained through 4\min.9 apertures yield the following global,
smoothed results: A(\ha )\,=\,0.15 mag, A(\ha\,-radio)\,=\,0.34
mag, A(1500\,\AA-FIR)\,=\,1.10 mag, and  
A(1900\,\AA-FIR)\,=\,0.66 mag \citep{caplan85,caplan86,bell02}.  \\

\subsection{Excitation}

\begin{figure*}
\begin{center}
\includegraphics[width = 8.5cm]{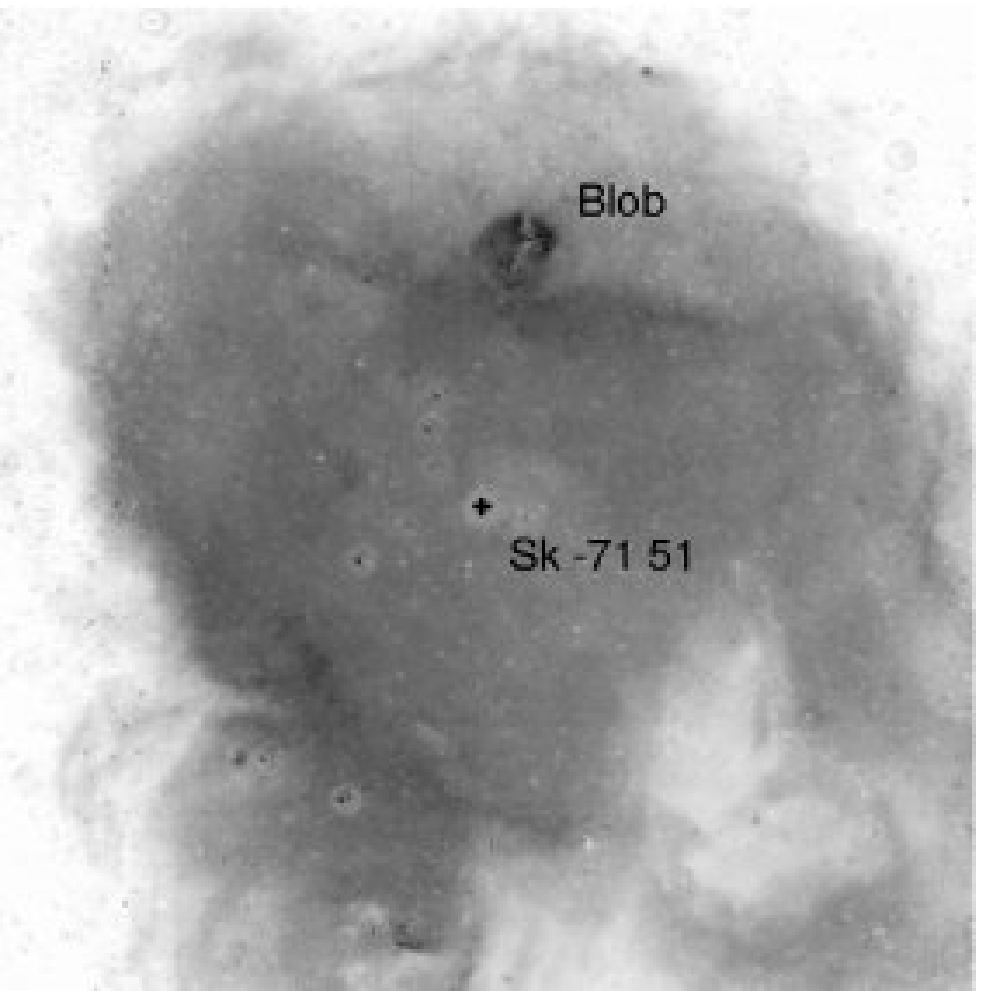}
\hfill
\includegraphics[width = 8.5cm]{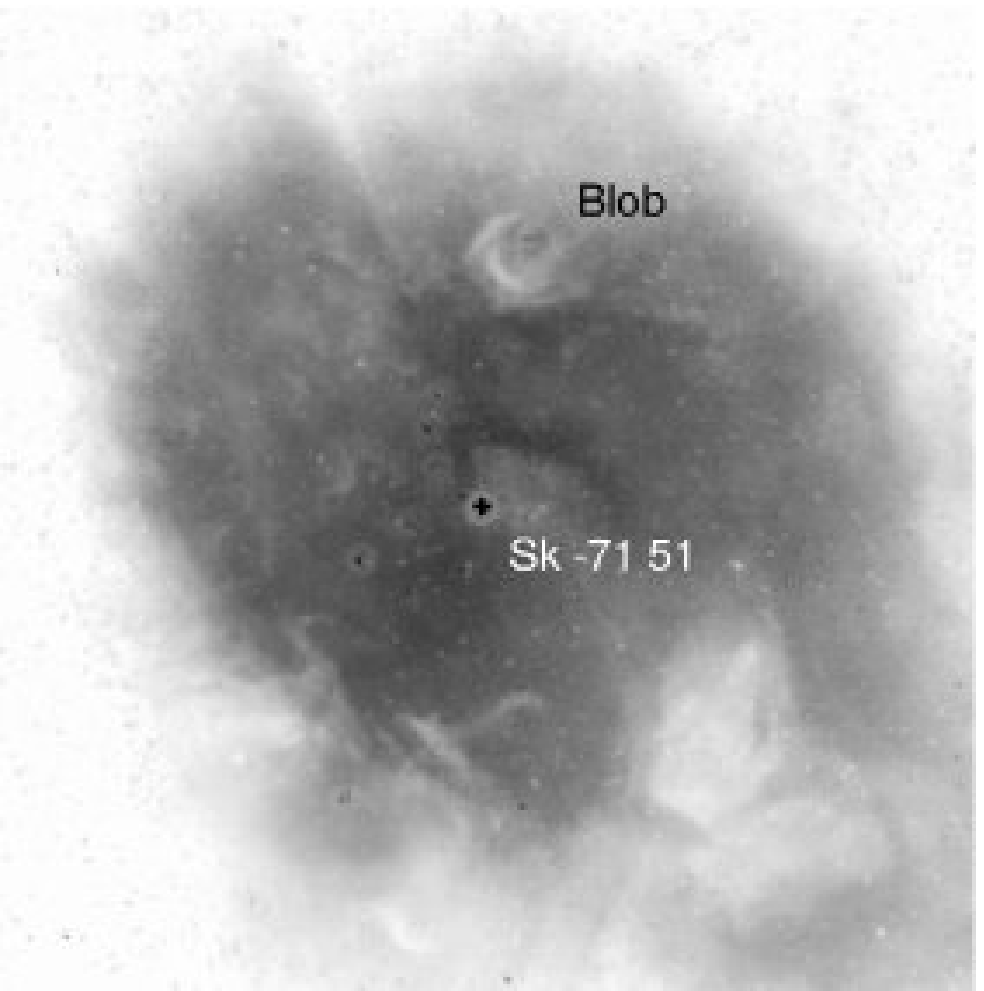}
\caption{Line ratio maps of the \h2 region N\,214C 
  based on  narrow-band filter images, 
after subtraction of point sources with \texttt{DAOPHOT/allstar}.  
The white circles around the stars are processing artefacts.
  {\bf Left:} The extinction \ha\,/\hb\, map.  
  {\bf Right:} The excitation  \oiii\,/\hb\, map. Field size as 
  in  Fig.\,\ref{fig:schema}.  North is up and
east to the left. 
\label{o3hb}  
}
\end{center}
\end{figure*}

The \oiii\,(5007\,\AA)/\hb\, ratio (Fig.\,\ref{o3hb}) 
has a uniform  value of about 5 
over an extended area centered on Sk\,$-71^{\circ}51$. In particular, the narrow nebular 
arc situated north of Sk\,$-71^{\circ}51$\, can be distinguished by its slightly  
higher value of \ab\,5.5. The ridge on which the blob seems to lie has 
a ratio between 4 and 5. \\

Regarding the ionized blob, the western lobe has a higher 
\oiii /\hb\, ratio attaining a value of 5.3 at most toward the 
bright emission area. This higher excitation zone is centered  
on star \#\,1145 but cut by the dust lane. The eastern 
lobe has a smaller mean value  of \ab\,3.5 and rises to 
\ab\,3.8 toward the north-eastern part of the lobe.

\subsection{Multiplicity of Sk\,--71$^{\circ}51$}

Although it was known that Sk\,$-71^{\circ}51$\, is not a single
star but a compact cluster \citep{garmany87}, its 
attributed luminosity was  based on low-resolution
observations obtained using a 61 cm telescope 
with an aperture 18\frac\, in size \citep{isser75}.
Apart from the fact that Sk\,$-71^{\circ}51$\, is a tight cluster, the presence  of 
a relatively bright star (\#\,9), detached from the main cluster but 
possibly present in the aperture, leads to an overstimation of the 
magnitude.  \\

The result of the image restoration by deconvolution, as explained 
in Sect. 2.2, for a 256$^{2}$ pixels field, corresponding to 
21\frac.8$^{2}$ on the sky, centered on Sk\,$-71^{\circ}51$, is presented in  
Fig.\,\ref{fig:deconvolution} and listed in  Table\,\ref{tab:deconvolution}
 which also gives the astrometry and photometry of the stars. 
The tight core of the Sk\,$-71^{\circ}51$\, cluster, 
covering a \ab\,4\,\frac\, area, is made up of at least 6 components, stars 
\#\,17, 14, 21, 19, 13, and 12. The brightest component, \#\,17   
with $V=12.85$,  $B-V=-0.15$, and $V-R=-0.06$ mag, 
is separated by \ab\,1\frac\, from the second 
brightest star, \#\,14 with $V=16.60$, $B-V=-0.09$, and $V-R=-0.06$ mag.    
Interestingly, the $V$ and $B-V$ magnitudes for star \#17\, agree well with 
\citet{oey}'s results. We notice that the 
present higher resolution data reduce the brightness of Sk\,$-71^{\circ}51$\, 
by 0.14 mag with respect to \citep{isser75}: $V=12.71$, $B-V=-0.09$, 
$U-B=-1.00$ mag.

\begin{figure*}
\centering\includegraphics[width=\linewidth]{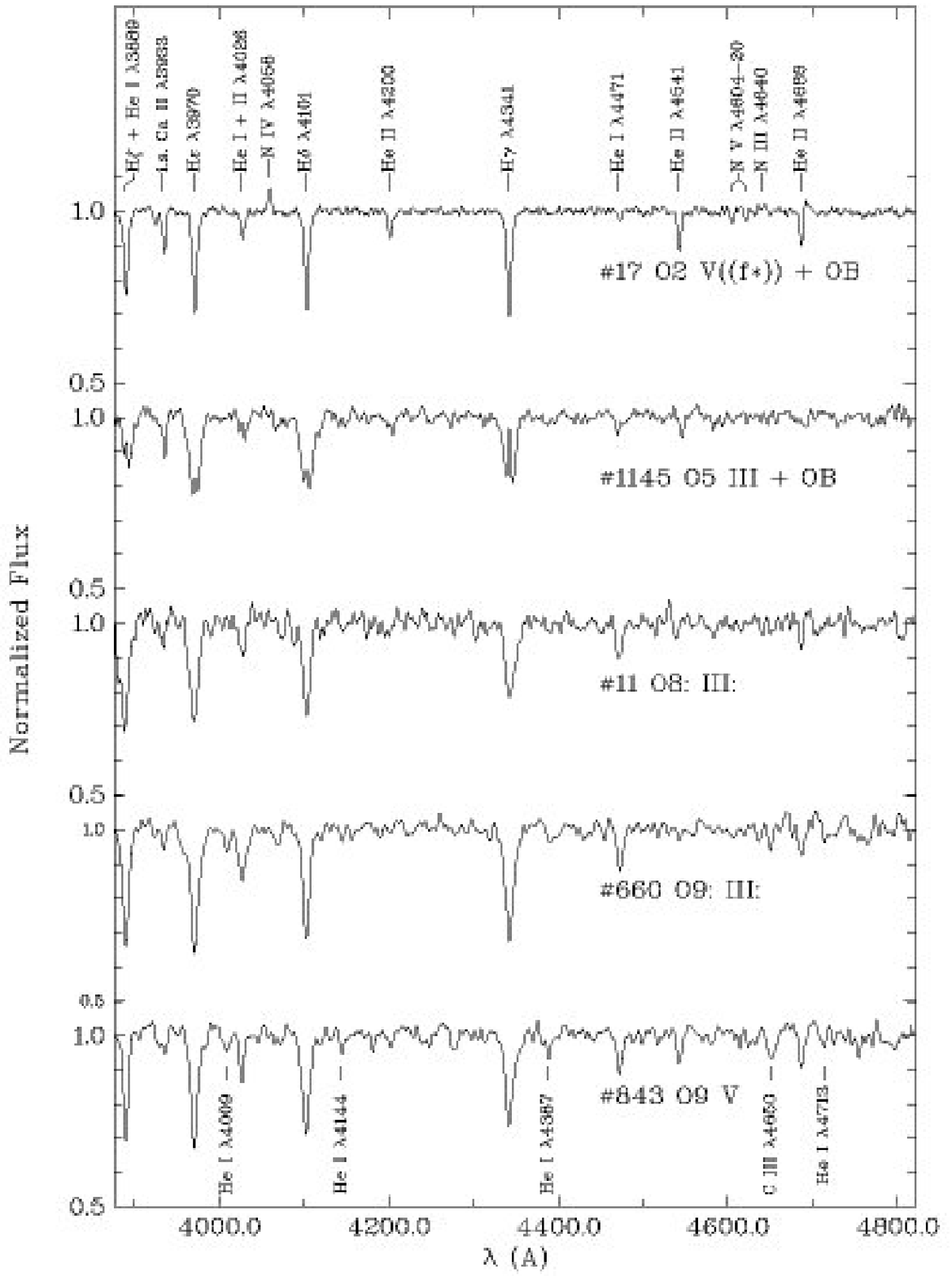}
\caption{Spectra of O type stars in the LMC \h2\, region N\,214C.}
\label{fig:otypes}
\end{figure*}

\begin{figure*}
\centering\includegraphics[width=\linewidth]{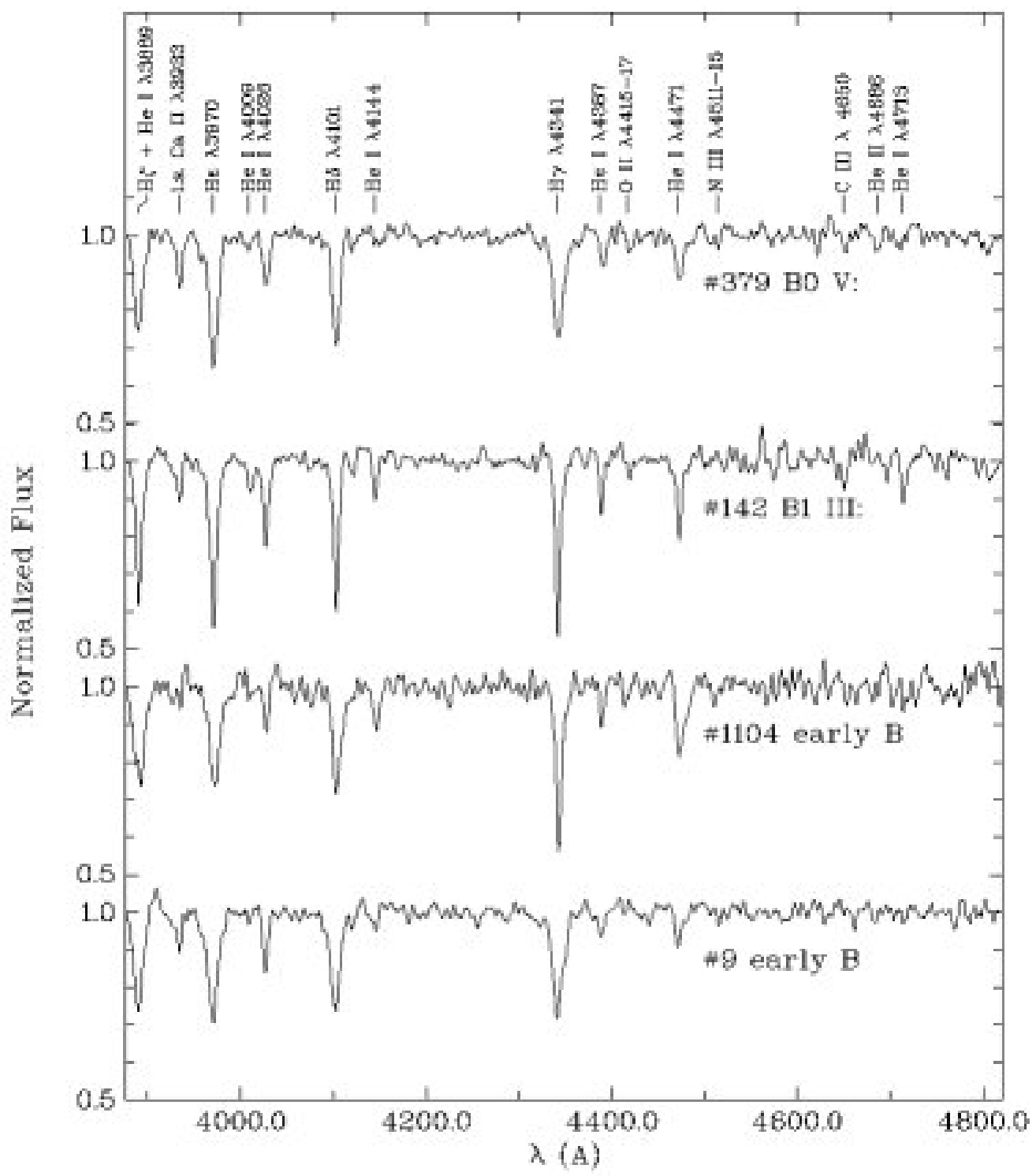}
\caption{ Spectra of B type stars in  the LMC \h2 region N\,214C. }
\label{fig:btypes}
\end{figure*}

\subsection{Spectral types}

The stars for which the spectroscopy was obtained are 
indicated in Fig.\,\ref{fig:schema}, while the  
spectrograms are displayed in Figs.\,\ref{fig:otypes} 
and \ref{fig:btypes}. The spectral 
classification was performed without knowledge of the stellar 
identifications or locations, with reference to the criteria 
and atlas of \citet{wal-fitz}.
The results are summarized in  Table\,\ref{tab:classification}  
which also gives the corresponding photometric and astrometric 
information. In the following some details of 
the two most interesting cases are given.

\subsubsection{Star Sk\,$-71^{\circ}51$ }

Star \#17 is the brightest component of the Sk\,$-71^{\circ}51$\, cluster 
(Table\,\ref{tab:deconvolution}). 
The spectral classification of Sk\,$-71^{\circ}51$\, has changed several
times as a function of the quality of the spectra used.  Initially it
was classified as an early type dwarf,  O4 V \citep{conti86}, but
subsequently was revised to an earlier giant O3 III(f*)
\citep{garmany87}. More recently, 
Sk\,$-71^{\circ}51$\, was  classified  as O3 V((f*)) by \citet{walborn02}. 
The present higher S/N ratio spectra indicate an even earlier 
main sequence star of type O2 V((f*)) due to a large 
\niv\,(\lam\,4058)/\niii\,(\lam\,4640) ratio. Note also the
presence of \heii\, \lam\,4686 P\,Cyg profile.  
The presence of the \hei\, 4471 line suggests an OB companion; 
it is quite possible that the \hei\, line arises in the close resolved 
companions, similarly to the case of LH10-3209 
\citep{walborn99}. However, contamination by  
unresolved close components can not be  excluded.  
Assuming an intrinsic color of $E(B-V)$\,=\,--0.32 mag for early O type   
stars \citep{walborn02} and a distance modulus of 18.6\,mag, the 
absolute magnitude of star \#\,17 is  M$_{V}$\,=\,--6.28.   \\

\subsubsection{Star \#\,1145}

Of spectral type  O5 III + OB, this is the brightest star 
($V=14.96$, $B-V=-0.06$ mag) apparently 
associated with the blob. The doubled line profiles in its spectrogram  
suggest a spectroscopic binary with    
$\delta$\,v\,\ab\,530 km s$^{-1}$.  
The companion dominates at \hei\,\lam\,4471 but the primary at
\hei\,+\,{\sc ii}\,\lam\,4026. \\

\begin{table*}
\setcounter{table}{3}
\caption[]{Spectral classication and photometry of  interesting stars 
    in LH\,110}
\begin{tabular}{ccccccccc}   
\hline
\hline
Star & $\alpha$ & $\delta$ &  $V$   & $B-V$   & $V-R$ & Spectral type &
 Notes \\
     &  (J2000) & (J2000)  & (mag)  & (mag) & (mag) &  &      &\\
\hline
 9 & 05:41:38.48 & --71:19:59.1 &   15.96 &   --0.14 &   --0.07 &early B &\\
11 & 05:41:37.60 & --71:19:58.6 &   15.70 &   --0.11 &   --0.06 &O8: III: &\\
17 & 05:41:39.62 & --71:19:56.4 &   12.85 &   --0.15 &   --0.06 &
   O2\,V((f*)) + OB & Main component of Sk\,$-71^{\circ}51$ \\
142 & 05:41:42.54 & --71:21:19.4 &   15.32 &   --0.12 &   --0.03 &B1 III: &\\
325 & 05:41:45.19 & -71:20:54.2 &   14.81 &   -0.14 &   -0.03 &\\
379 & 05:41:48.37 & --71:20:46.1 &   15.40 &   --0.11 &   --0.03 &
B0 V: &\\
437 & 05:41:26.14 & -71:20:37.8 &   16.07 &    2.19 &    1.22 &  \\
660 & 05:41:44.53 & --71:20:06.8 &   14.92 &   --0.15 &   --0.03 & O9: III: &\\
709 & 05:41:42.29 & --71:19:59.7 &   15.85 &    0.83 &    0.55 &late-type&\\
843 & 05:41:41.70 & --71:19:40.6 &   15.53 &   --0.16 &   --0.04 &O9 V&\\
1104 & 05:41:59.18 & --71:19:20.4 &   20.01 &    0.37 &    0.18 &
          early B &  Associated with the Blob      \\
1121 & 05:41:36.51 & -71:19:05.8 &   16.91 &    0.03 &    0.12 & &
Seen toward the western lobe\\
1132 & 05:41:38.16 &  -71:19:04.3 &   19.41 &     0.05 &    0.12 &  &
          Seen toward the center of the Blob \\
1136 & 05:41:39.30 & -71:19:03.9 &   16.37 &    1.66 &    0.98 &&
Seen toward the eastern lobe\\
1145 & 05:41:37.68 & --71:19:03.2 &   14.96 &   --0.06 &    0.04 &O5\,III
   + OB & Associated with the Blob\\
1846 & 05:41:23.32 & -71:17:32.8 &   18.19 &    0.99 &    0.56 & & 
   Associated with N\,214H-3 \\
1864 & 05:41:22.89 & -71:17:29.8 &   15.68 &   -0.00 &    0.01 & &
    Associated with N\,214H-3 \\
1882 & 05:41:25.43 & -71:17:27.2 &   15.03 &    0.71 &    0.41 & &
   See Fig.\,\ref{fig:anonymous} \\
1936 & 05:41:22.58 & -71:17:19.2 &   17.28 &    0.22 &    0.18 & &
    Associated with N\,214H-2\\
1964 & 05:41:26.39 & -71:17:14.8 &   16.81 &    0.08 &    0.13 & &
    Associated with N\,214H-4\\
2065 & 05:41:26.61 & -71:16:59.5 &   14.84 &   -0.10 &   -0.06 & &
    Associated with N\,214H-4\\
\hline
\label{tab:classification}
\end{tabular}
\end{table*}

\section{Discussion}

\subsection{Stellar content}

The \h2 region N\,214C lies in  the OB association LH\,110 
of size 3\min\,\,\x\,2\min\,\, \citep{lh}, in which the latter 
authors detected seven OB candidates. 
The color-magnitude diagram obtained for 2341 stars of the OB association  
is displayed in Fig.\,\ref{col-mag}. Overplotted are also  
the isochrones with ages 1 Myr, 1 Gyr, and 10 Gyr 
for  metallicity Z = 0.008 obtained by \citet{lejeune01}.  
Two main stellar populations show up in this diagram: a main sequence 
and an evolved component. The main sequence, well fitted by 
an isochrones of age 1 Myr with a reddening of $A_V = 0.5$\,mag, 
is made up  of stars 
with initial masses from \ab\,80\,\sm\, to \ab\,4\,\sm\, and possibly 
as low as \ab\,2\,\sm. The color spread for the lower mass main 
sequence stars may be due to either  the reddening effect or 
contamination by the evolved stars. \\

The evolved population consists of stars which are well fitted 
by a 1 Gyr isochrone, corresponding to a turnoff mass between 
1.9 and 2.0\,\sm. Stars less massive than the turnoff mass have 
$(B-V)$ colors between 0.0 and 0.5 mag, whereas more massive stars are 
centered on $(B-V)=$\,\ab\,1 mag. 
Note that the 10 Gyr track is tangent to the redder border of the 
evolved population, probably indicating that the bulk of the 
stars are better fitted by a much younger, 1 Gyr isochrones. \\ 

Star \#\,709 is evolved, as indicated by its spectrum presenting features 
typical of evolved low-mass stars. This result is supported by the 
position of the star on the color-magnitude diagram. 
Its absolute magnitude, $-2.7$, would point to a late-type  
M giant. This star is quite  interesting since it lies on the face of 
the \h2 region N\,214C not far from the Sk\,$-71^{\circ}51$\, cluster.
Generally speaking, several stars seen directly toward N\,214C 
belong to the evolved population. Do they really pertain to this 
region, or are they field stars? Radial velocity determinations 
are necessary to answer this question. Another remarkable case is 
star \#\,437, which has the reddest colors of the sample 
(see also Table\,\ref{tab:classification}). It lies toward the 
loop structure of N\,214C (Sect. 3.1) where the reddening, while not 
negligible, is not enough to explain the strong colors measured 
for \#\,437. Comparison with the intrinsic $B-V$ and $V-R$ colors 
\citep{houdashelt00} suggests an evolved low-mass, field star 
likely affected by local extinction. \\

The fact that the \h2 blob lies toward the northern 
outskirts of N\,214C and the compact \h2 regions N\,214H-1, 2, and 3 
are situated in that area, suggests that the detected molecular cloud 
\citep{israel93, heikkila} is in contact with the ionized gas in 
that direction. This assumption 
agrees with the observation that stars \#\,1882 and \#\,1846 have a 
red color (Fig.\,\ref{col-mag}). Note also that stars 
\#\,1936 and more especially \#\,1964, the latter situated 
in the diffuse, and maybe older \h2 region N\,214H-4, have a smaller 
reddening, because they may be less affected by the molecular cloud. \\

The upper part of the main sequence is populated by massive O and B 
types  several of which have been classified in this work.  
These stars, seen on the face of the \h2 region N\,214C, are associated 
with the \h2 region and therefore contribute to its ionization. 
Their initial masses are $<$ 40\,\sm, with a very remarkable exception: 
star \#\,17 with a mass of \ab\,80\,\sm\, if single. This main component 
of the Sk\,$-71^{\circ}51$\, cluster is a very hot star of type O2 V((f*)), recently 
introduced by \citet{walborn02}. This type is very rare since so far only 
a dozen members  have been identified. Should star \#\,17 
not be fully resolved, as implied by the presence of  \hei\,\lam\,4471 
in its spectrum, its mass must be smaller than 80\,\sm. 
Note however that the later-type OB signature may be due to stars \#\,14 
or \#\,21 lying 1\frac.2 and 1\frac.4 respectively from \#\,17 and not 
to a binary companion. In that case, the estimated mass 
of \ab\,80\,\sm\, based only on photometry, will be maintained. 
 Radial velocity observations may elucidate   
the situation. Anyhow, star \#\,17, as the main excitation source of 
N\,214C, should have a powerful wind, as revealed by the presence of 
wind and shock  features created in the nebula around the star 
(Sect. 3.1). \\

The blue star \#\,325 associated with the south-eastern extension of 
N\,214C must be an O type of initial mass 30--40\,\sm, as indicated 
by the color-mag diagram. Although we do not have spectroscopic 
observations, the fact that the star is surrounded by \oiii\, emission 
in a region  globally dominated by \ha\, emission supports   
an O type characteristic. Note that stars \#\,379 and \#\,142, 
which are classified B0 V: and B1 III: (Fig.\,\ref{fig:btypes}, 
Table\,\ref{tab:classification}  and lie toward the same 
region as \#\,325, are not associated with \oiii\, emission 
due to the lack of sufficiently hard ionizing photons in B stars.

\subsection{The peculiar Blob}

The compact \h2 nebula located \ab\,60\frac\, north of Sk\,$-71^{\circ}51$\, has 
a striking spherical shape of \ab\,5\frac\, (\ab\,1.3 pc) in radius.
Its  mass would be \ab\,100\,\sm, assuming a sphere of 1\,pc in radius 
filled with atomic hydrogen of  density 1000 cm$^{-3}$.
The excitation of the blob may be mainly due to star \#\,1145, 
type O5 III + OB,  which probably lies outside the sphere. 
The fact that the bright emission peak, where the  
\oiii/\hb\, ratio has its highest value, lies close 
to \#\,1145 confirms the exciting role of this star. This result is also in 
line with high-resolution radio continuum observations at 3 cm (8.6 GHz) 
and 6 cm (4.8 GHz) obtained using the Australia Telescope Compact Array 
\citep{indebetouw04}. These observations indicate the presence of a 
compact, ionized region (\object{B0542-7121}) at the position of the blob with 
radio continuum fluxes $19 \pm 2$ and $18 \pm 1.5$ mJy at 3 and 6 cm 
respectively. These measurements correspond to a Lyman continuum flux 
of 4.80\,\x\,10$^{48}$ photons s$^{-1}$ assuming that the \h2 region 
is ionization bounded. The exciting star needed to provide this flux 
is about O9 V type \citep{vacca96}. 
The discrepancy between this estimate and the spectral classification  
(O5 III star) can be explained by 
the fact that the star is outside the blob, so only a fraction 
of its radiation intercepts in.
In case an ionizing source 
is hidden in the blob the extinction should be so strong that no Lyman 
continuum photons can escape. 
The radio continuum observations confirm also that 
star \#\,1104, an early B, does not have a major role in the 
ionization of the blob. On the other hand,  there is a faint star 
\#\,1132 seen toward the center of the blob in the dark opening, 
but this is most likely a foreground, blue star of $V=19.41, 
B-V = 0.05$ mag.  \\

The fact that no hidden exciting source is expected does not preclude
the presence of an internal infrared object. And, interestingly, a
strong IRAS source, \object{05423-7120}, coincides with the blob 
\citep{indebetouw04}. 
The measured fluxes ($F_{12}$\,=\,0.84 Jy, $F_{25}$\,=\,3.91 Jy, 
$F_{60}$\,=\,45.74 Jy, and  $F_{100} <$ 110.80 Jy) correspond to a 
luminosity of \ab\,2\,\x\,10$^{5}$ \slum\, as
derived from the 60\,$\mu$m flux. This estimate implies a massive heat
source, equivalent to an embedded \ab\,O7~V star of mass \ab\,40\,\sm. 
The IRAS data can be fitted by a blackbody with $T=60$ K whose maximum 
is centerd on \ab\,60\,$\mu$m.
This suggests that there is indeed an embedded massive star not
just a cold collapsing gas clump whose spectral energy distribution
would probably peak at or beyond 100\,$\mu$m. 
There is however an excess emission at 12\,$\mu$m. Although the 
excess in some cases could be due to a component of very small 
grains or polycyclic aromatic hydrocarbons \citep{degioia92, bell02},   
it could also be due to the presence of a hotter source 
\citep{wolf-chase03}.  The latter suggestion is in line with 
the expectation that the putative object is most likely not isolated  but 
embedded in an infrared cluster.
Alternatively, an accreting
protostar cannot be ruled out.  In that case a very massive protostar
(\ab\,100\,\sm\,) with a high accretion rate 
(\ab\,10$^{-3}$\,\sm\,yr$^{-1}$) is
required. Another possibility would be a trapped ultracompact \h2 
region created immediately following high mass star formation \citep{keto02}. 
This would imply a low radio luminosity compared to
the total luminosity reradiated by dust, suggesting also that the
object might be so heavily obscured that it emits mainly in far
infrared. If this putative object exists, it would be interesting to
determine its near- and mid-infrared properties using high spatial
resolution observations. \\

The compact \h2 region discovered in N\,214C may be a newcomer 
to the family of HEBs (High Excitation Blobs) 
in the Magellanic Clouds, the first member of which was detected in 
LMC N\,159 (\citet{mhm99c} and references therein). In contrast to the typical 
\h2 regions of the Magellanic Clouds, which are extended structures 
spanning several minutes of arc on the sky (more than 50 pc) and powered by a
large number of hot stars, HEBs are dense, small regions usually
5\frac\, to 10\frac\, in diameter (1 to 3 pc). Moreover, they often
form adjacent to or apparently inside the typical giant \h2 regions, 
and rarely
in isolation. They are generally affected by significant amounts 
of local dust. The formation mechanisms of these objects are not yet 
well understood, in the sense that we do not know which particular 
conditions can give rise to them adjacent to typical Magellanic 
Cloud \h2 regions. One thing seems however sure, they represent the youngest 
massive stars of their OB associations. So far only a half-dozen 
of them have been detected (SMC N\,81, N88A; LMC N\,159, N\,83B, N\,11A, 
N\,160A1 \& A2) and studied using the Hubble Space  Telescope 
\citep{mhm99a,mhm99b,mhm99c,mhm01a,mhm01b,mhm02a,mhm02b}. 
But the exciting stars of the tightest or youngest 
members of the family remain undetected even with the {\it HST} 
angular resolution in the visible. \\

The unusual spherical shape of the blob may be due to lack of 
angular resolution, if we compare it with the Galactic \h2 region 
NGC\,2024 (Flame nebula), which has a linear radius of 
\ab\,1.5 pc, 
comparable to the \h2 blob. 
The optical image of NGC\,2024 also shows a central elongated obscuration 
in the north-south direction (see \citet{lenorzer04} and references 
therein). There is a B-type infrared cluster behind the dust lane, and  
recently \citet{bik03} estimated a spectral type of \ab\,O8 
for the dominant source of ionizing flux for the \h2 region.  
In spite of these apparent similarities, 
two important differences distinguish the N\,214C blob from NGC\,2024. 
The most massive star of the latter is an O8 type accompanied by 
a cluster of B types, while the blob is associated with a more massive 
star of type O5 III + OB (\#\,1145). Moreover, the blob lies in a 
prominent region which has formed several O types and more especially 
a very rare, massive, and hot star of type O2 V. These characteristics 
call for high-resolution infrared observations of the blob in order 
to investigate deep into the blob, specifically behind the absorption 
lane. It would be very interesting if the embedded stellar population of 
the blob turns out to be similar to that of NGC\,2024. This will imply 
that sharply different environmental factors (for instance metallicity 
and initial mass function) can bring about similar stellar populations 
under certain conditions. Knowing these particular conditions will be 
important for better understanding star and cluster formation. \\
  
The ridge feature upon which the \h2 blob seems to lie, may 
in fact be an ionization front moving northward into the molecular cloud. 
If this assumption is right, the  blob may have resulted from  
massive star formation following the collapse of a thin shell of neutral 
matter accumulated between the shock and ionization fronts, as 
predicted by the sequential star formation scenario \citep{elmegreen77}. 
A list of carefully selected Galactic candidate regions which are likely 
to be examples of this star formation process is  presented by 
\citet{deharveng04}. Based on {\it HST} observations, this scenario  
has also been 
suggested for LMC N\,83B \citep{mhm01a}. Since star \#\,1145 and 
the blob lie apparently close to each other, one can raise the question 
whether that star also is triggered by the ionization front of star \#\,17. 
Star \#\,1145 (O5 III + OB) is less massive than \#\,17 (O2 V + OB) and 
should naturally evolve more slowly than the latter. Yet the luminosity class 
of \#\,1145 seems to imply the contrary. It should be underlined that 
the luminosity class III means that the absorption feature  
\hei\,\,\lam\,4686 is filled in by emission.  But this description of the 
spectrum  could have different physical causes.  Usually 
that is due to higher luminosity, but anything that produces the 
\hei\,\,\lam\,4686 emission will fill it in, e.g. loose material/colliding 
winds in the binary. Therefore, one cannot rule out the possibility that 
\#\,1145 be a second generation star triggered by \#\,17. \\

Fig.\,\ref{fig:deep} reveals another ridge southeast of N\,214C, 
suggesting that the \h2 region is in contact with a molecular cloud also 
in that direction. Since the massive stars \#\,142, \#\,325, and \#\,379 
are apparently situated beyond the south-eastern ridge, 
the question of sequential star formation can  similarly be  
raised for them.  \\

\acknowledgements{We are grateful to Drs. P. Magain, F. Courbin and their 
  team for giving us access to their new deconvolution code and for their
  hospitality at their Institute in Li\`ege.  
  We are also indebted to Dr. J.\,Walsh (ST-ECF) for his tool, 
  \texttt{specres}, and the useful 
  advices about its use. We are also thankful to Dr. L.\,Germany 
  (ESO, NTT) for providing valuable customized calibration data, which 
  greatly improved our data reduction. We are also grateful to 
  Drs. Lise Deharveng (Laboratoire d'Astrophysique de Marseille), 
  Thibaut Le Bertre (LERMA, Paris Observatory), and Hans Zinnecker 
  (Astrophysikalisches Institut Potsdam) for 
  discussions. We would like also to thank the referee, Dr. Paul Crowther, 
  for helpful remarks and comments.    
}

\bibliographystyle{aa}
\bibliography{meynadierbib}

\def\apjs{ApJS} \def\apj{ApJ} \def\apjl{ApJL} \def\aap{A\&A} \def\aaps{A\&AS}
  \def\aj{AJ} \def\pasp{PASP} \def\mnras{MNRAS} \def\aapr{AARev}
\begin{thebibliography}{41}
\expandafter\ifx\csname natexlab\endcsname\relax\def\natexlab#1{#1}\fi

\bibitem[{{Bell} {et~al.}(2002){Bell}, {Gordon}, {Kennicutt}, \&
  {Zaritsky}}]{bell02}
{Bell}, E.~F., {Gordon}, K.~D., {Kennicutt}, R.~C., \& {Zaritsky}, D. 2002,
  \apj, 565, 994

\bibitem[{{Bertin} \& {Arnouts}(1996)}]{bertin96}
{Bertin}, E. \& {Arnouts}, S. 1996, \aaps, 117, 393

\bibitem[{{Bik} {et~al.}(2003){Bik}, {Lenorzer}, {Kaper}, {Comer{\' o}n},
  {Waters}, {de Koter}, \& {Hanson}}]{bik03}
{Bik}, A., {Lenorzer}, A., {Kaper}, L., {et~al.} 2003, \aap, 404, 249

\bibitem[{{Caplan} \& {Deharveng}(1985)}]{caplan85}
{Caplan}, J. \& {Deharveng}, L. 1985, \aaps, 62, 63

\bibitem[{{Caplan} \& {Deharveng}(1986)}]{caplan86}
{Caplan}, J. \& {Deharveng}, L. 1986, \aap, 155, 297

\bibitem[{{Chin} {et~al.}(1997){Chin}, {Henkel}, {Whiteoak}, {Millar}, {Hunt},
  \& {Lemme}}]{chin}
{Chin}, Y.-N., {Henkel}, C., {Whiteoak}, J.~B., {et~al.} 1997, \aap, 317, 548

\bibitem[{{Conti} {et~al.}(1986){Conti}, {Garmany}, \& {Massey}}]{conti86}
{Conti}, P.~S., {Garmany}, C.~D., \& {Massey}, P. 1986, \aj, 92, 48

\bibitem[{Davies {et~al.}(1976)Davies, Eliott, \& Meaburn}]{dem}
Davies, R., Eliott, K., \& Meaburn, J. 1976, Mem. R. Astron. Soc., 81, 89

\bibitem[{{Degioia-Eastwood}(1992)}]{degioia92}
{Degioia-Eastwood}, K. 1992, \apj, 397, 542

\bibitem[{{Deharveng} {et~al.}(2005){Deharveng}, {Zavagno}, \&
  {Caplan}}]{deharveng04}
{Deharveng}, L., {Zavagno}, A., \& {Caplan}, J. 2005, \aap, 433, 565

\bibitem[{{Elmegreen} \& {Lada}(1977)}]{elmegreen77}
{Elmegreen}, B.~G. \& {Lada}, C.~J. 1977, \apj, 214, 725

\bibitem[{{Garmany} \& {Walborn}(1987)}]{garmany87}
{Garmany}, C.~D. \& {Walborn}, N.~R. 1987, \pasp, 99, 240

\bibitem[{{Groenewegen} \& {Oudmaijer}(2000)}]{groenewegen00}
{Groenewegen}, M.~A.~T. \& {Oudmaijer}, R.~D. 2000, \aap, 356, 849

\bibitem[{{Heikkila} {et~al.}(1998){Heikkila}, {Johansson}, \&
  {Olofsson}}]{heikkila}
{Heikkila}, A., {Johansson}, L.~E.~B., \& {Olofsson}, H. 1998, \aap, 332, 493

\bibitem[{{Henize}(1956)}]{henize}
{Henize}, K.~G. 1956, \apjs, 2, 315

\bibitem[{{Heydari-Malayeri} {et~al.}(2002{\natexlab{a}}){Heydari-Malayeri},
  {Charmandaris}, {Deharveng}, {Meynadier}, {Rosa}, {Schaerer}, \&
  {Zinnecker}}]{mhm02a}
{Heydari-Malayeri}, M., {Charmandaris}, V., {Deharveng}, L., {et~al.}
  2002{\natexlab{a}}, \aap, 381, 941

\bibitem[{{Heydari-Malayeri} {et~al.}(2001{\natexlab{a}}){Heydari-Malayeri},
  {Charmandaris}, {Deharveng}, {Rosa}, {Schaerer}, \& {Zinnecker}}]{mhm01b}
{Heydari-Malayeri}, M., {Charmandaris}, V., {Deharveng}, L., {et~al.}
  2001{\natexlab{a}}, \aap, 372, 527

\bibitem[{{Heydari-Malayeri} {et~al.}(2001{\natexlab{b}}){Heydari-Malayeri},
  {Charmandaris}, {Deharveng}, {Rosa}, {Schaerer}, \& {Zinnecker}}]{mhm01a}
{Heydari-Malayeri}, M., {Charmandaris}, V., {Deharveng}, L., {et~al.}
  2001{\natexlab{b}}, \aap, 372, 495

\bibitem[{{Heydari-Malayeri} {et~al.}(1999{\natexlab{a}}){Heydari-Malayeri},
  {Charmandaris}, {Deharveng}, {Rosa}, \& {Zinnecker}}]{mhm99b}
{Heydari-Malayeri}, M., {Charmandaris}, V., {Deharveng}, L., {Rosa}, M.~R., \&
  {Zinnecker}, H. 1999{\natexlab{a}}, \aap, 347, 841

\bibitem[{{Heydari-Malayeri} {et~al.}(2003){Heydari-Malayeri}, {Meynadier}, \&
  {Walborn}}]{hmmw03}
{Heydari-Malayeri}, M., {Meynadier}, F., \& {Walborn}, N.~R. 2003, \aap, 400,
  923

\bibitem[{{Heydari-Malayeri} {et~al.}(1999{\natexlab{b}}){Heydari-Malayeri},
  {Rosa}, {Charmandaris}, {Deharveng}, \& {Zinnecker}}]{mhm99c}
{Heydari-Malayeri}, M., {Rosa}, M.~R., {Charmandaris}, V., {Deharveng}, L., \&
  {Zinnecker}, H. 1999{\natexlab{b}}, \aap, 352, 665

\bibitem[{{Heydari-Malayeri} {et~al.}(2002{\natexlab{b}}){Heydari-Malayeri},
  {Rosa}, {Schaerer}, {Martins}, \& {Charmandaris}}]{mhm02b}
{Heydari-Malayeri}, M., {Rosa}, M.~R., {Schaerer}, D., {Martins}, F., \&
  {Charmandaris}, V. 2002{\natexlab{b}}, \aap, 381, 951

\bibitem[{{Heydari-Malayeri} {et~al.}(1999{\natexlab{c}}){Heydari-Malayeri},
  {Rosa}, {Zinnecker}, {Deharveng}, \& {Charmandaris}}]{mhm99a}
{Heydari-Malayeri}, M., {Rosa}, M.~R., {Zinnecker}, H., {Deharveng}, L., \&
  {Charmandaris}, V. 1999{\natexlab{c}}, \aap, 344, 848

\bibitem[{{Houdashelt} {et~al.}(2000){Houdashelt}, {Bell}, {Sweigart}, \&
  {Wing}}]{houdashelt00}
{Houdashelt}, M.~L., {Bell}, R.~A., {Sweigart}, A.~V., \& {Wing}, R.~F. 2000,
  \aj, 119, 1424

\bibitem[{{Indebetouw} {et~al.}(2004){Indebetouw}, {Johnson}, \&
  {Conti}}]{indebetouw04}
{Indebetouw}, R., {Johnson}, K.~E., \& {Conti}, P. 2004, \aj, 128, 2206

\bibitem[{{Israel}(1997)}]{israel97}
{Israel}, F.~P. 1997, \aap, 328, 471

\bibitem[{{Israel} {et~al.}(1993){Israel}, {Johansson}, {Lequeux}, {Booth},
  {Nyman}, {Crane}, {Rubio}, {de Graauw}, {Kutner}, {Gredel}, {Boulanger},
  {Garay}, \& {Westerlund}}]{israel93}
{Israel}, F.~P., {Johansson}, L.~E.~B., {Lequeux}, J., {et~al.} 1993, \aap,
  276, 25

\bibitem[{{Isserstedt}(1975)}]{isser75}
{Isserstedt}, J. 1975, \aaps, 19, 259

\bibitem[{{Keto}(2002)}]{keto02}
{Keto}, E. 2002, \apj, 580, 980

\bibitem[{{Lejeune} \& {Schaerer}(2001)}]{lejeune01}
{Lejeune}, T. \& {Schaerer}, D. 2001, \aap, 366, 538

\bibitem[{{Lenorzer} {et~al.}(2004){Lenorzer}, {Bik}, {de Koter}, {Kurtz},
  {Waters}, {Kaper}, {Jones}, \& {Geballe}}]{lenorzer04}
{Lenorzer}, A., {Bik}, A., {de Koter}, A., {et~al.} 2004, \aap, 414, 245

\bibitem[{{Lucke} \& {Hodge}(1970)}]{lh}
{Lucke}, P.~B. \& {Hodge}, P.~W. 1970, \aj, 75, 171

\bibitem[{{Lucy} \& {Walsh}(2003)}]{specres}
{Lucy}, L.~B. \& {Walsh}, J.~R. 2003, \aj, 125, 2266

\bibitem[{{Magain} {et~al.}(1998){Magain}, {Courbin}, \& {Sohy}}]{Magain98}
{Magain}, P., {Courbin}, F., \& {Sohy}, S. 1998, \apj, 494, 472

\bibitem[{{Magain} {et~al.}(2005){Magain}, {Courbin}, {Sohy}, {Gillon}, \&
  {Letawe}}]{magain04}
{Magain}, P., {Courbin}, F., {Sohy}, S., {Gillon}, M., \& {Letawe}, G. 2005, in
  prep.

\bibitem[{{Oey}(1996)}]{oey}
{Oey}, M.~S. 1996, \apjs, 104, 71

\bibitem[{{Vacca} {et~al.}(1996){Vacca}, {Garmany}, \& {Shull}}]{vacca96}
{Vacca}, W.~D., {Garmany}, C.~D., \& {Shull}, J.~M. 1996, \apj, 460, 914

\bibitem[{{Walborn} {et~al.}(1999){Walborn}, {Drissen}, {Parker}, {Saha},
  {MacKenty}, \& {White}}]{walborn99}
{Walborn}, N.~R., {Drissen}, L., {Parker}, J.~W., {et~al.} 1999, \aj, 118, 1684

\bibitem[{{Walborn} \& {Fitzpatrick}(1990)}]{wal-fitz}
{Walborn}, N.~R. \& {Fitzpatrick}, E.~L. 1990, \pasp, 102, 379

\bibitem[{{Walborn} {et~al.}(2002){Walborn}, {Howarth}, {Lennon}, {Massey},
  {Oey}, {Moffat}, {Skalkowski}, {Morrell}, {Drissen}, \& {Parker}}]{walborn02}
{Walborn}, N.~R., {Howarth}, I.~D., {Lennon}, D.~J., {et~al.} 2002, \aj, 123,
  2754

\bibitem[{{Wolf-Chase} {et~al.}(2003){Wolf-Chase}, {Moriarty-Schieven}, {Fich},
  \& {Barsony}}]{wolf-chase03}
{Wolf-Chase}, G., {Moriarty-Schieven}, G., {Fich}, M., \& {Barsony}, M. 2003,
  \mnras, 344, 809

\end{thebibliography}

\end{document}